\documentclass[aps,prl,reprint]{revtex4-1}

\usepackage{graphicx}
\usepackage{amsmath}

\usepackage{dsfont}
\newcommand*{\Z}{\mathds{Z}}

\begin{document}

\title{Many-body delocalization via symmetry emergence}

\author{N. S. Srivatsa}
\affiliation{Max-Planck-Institut f\"{u}r Physik komplexer Systeme, N{\"o}thnitzer Str.\ 38, D-01187 Dresden, Germany}
\author{Roderich Moessner}
\affiliation{Max-Planck-Institut f\"{u}r Physik komplexer Systeme, N{\"o}thnitzer Str.\ 38, D-01187 Dresden, Germany}
\author{Anne E. B. Nielsen}
\altaffiliation{On leave from Department of Physics and Astronomy, Aarhus University, DK-8000 Aarhus C, Denmark.}
\affiliation{Max-Planck-Institut f\"{u}r Physik komplexer Systeme, N{\"o}thnitzer Str.\ 38, D-01187 Dresden, Germany}

\begin{abstract}
Many-body localization (MBL) provides a mechanism to avoid thermalization in many-body quantum systems. Here, we show that an {\it emergent} symmetry can  protect a state from MBL. Specifically, we propose a $\Z_2$ symmetric model with nonlocal interactions, which has an analytically known, SU(2) invariant, critical ground state. At large disorder strength all states at finite energy density are in a glassy MBL phase, while the lowest energy states are not. These do, however, localize when a perturbation destroys the emergent SU(2) symmetry. The model also provides an example of MBL in the presence of nonlocal, disordered interactions that are more structured than a power law. The presented ideas raise the possibility of an `inverted quantum scar', in which a state that does not exhibit area law entanglement is embedded in an MBL spectrum, which does.
\end{abstract}

\maketitle

The eigenstate thermalization hypothesis suggests that clean quantum systems typically thermalize \cite{Deu,Sred}. Alternatively, disorder may prevent thermalization and instead give rise to MBL \cite{Basko}. This opens the door for generating interesting phases, in which e.g.\ excited states exhibit entanglement properties similar to ground states \cite{revNandkishore,Parameswaran_2018}. Examples of MBL have been found numerically in simple spin models \cite{Basko,revNandkishore} and studied in experiment \cite{MBLexp}, and many aspects of MBL are currently under investigation.

One area of intense interest is the interplay between symmetry and MBL. For instance, while it is known that MBL can be present in systems where the Hamiltonian has $\Z_2$ symmetry \cite{kjall}, it has been argued that short-range Hamiltonians with SU(2) symmetry cannot support an MBL phase, because the eigenstates of such Hamiltonians do not have area law entanglement \cite{Protopopov}. Also, it was noted that symmetry-constrained dynamics can yield a many-body mobility edge \cite{Mondragon}.

Another area of intense interest is the presence or absence of MBL in models with nonlocal interactions. Recent studies, which have focused on power law interactions and/or hopping in a random potential or random magnetic field, suggest that MBL can occur in long-range models \cite{MBLdipolar,Burin,Nand,Nag}. MBL has also been studied in models with power law interactions with random strengths or signs combined with a random magnetic field \cite{Tikhonov,Botzung}.

Here, we introduce a new type of model which exhibits MBL with a number of novel properties. The main idea is to use emergent symmetry of a single state in the spectrum to protect it from MBL, while not preventing the rest of the spectrum from localizing. Specifically, we investigate a Hamiltonian which has only a $\Z_2$ symmetry, but nevertheless has a critical ground state with SU(2) symmetry. We introduce disorder into this model via random positions of the spins, and we show that all states at finite energy density form a glassy phase \cite{order,kjall,Parameswaran_2018} at an appropriate disorder strength, while the ground state remains critical.

The model exhibits several interesting features. First, it shows that an emergent symmetry can have interesting and nontrivial effects with respect to MBL. Specifically, for a wide range of disorder strengths all the states at finite energy density form a MBL glass, while a few states at the bottom of the spectrum do not. Crucially, these states also become glassy as the emergent SU(2) symmetry vanishes upon perturbing the Hamiltonian. Second, it gives an example of MBL in a system with nonlocal, disordered interactions that have more structure than a power law. Third, the ground state can be found analytically, which allows for a detailed study of its properties even for large system sizes. Fourth, the model shows as yet unexplained gaps in the disorder averaged energy spectrum.

\textit{Model}--- We study a system of $N$ spin-$1/2$ particles on a unit circle and express the Hamiltonian
\begin{equation}\label{ham}
H=\sum_{i\neq j}F^{A}_{ij}(S_i^xS_j^x+S_i^yS_j^y)+\sum_{i\neq j}F_{ij}^{B}S^z_iS^z_j+F^{C}
\end{equation}
in terms of the spin operators $S_i^{a}=\sigma_i^a/2$, $a\in\{x,y,z\}$, where $\sigma_i^a$ are the Pauli matrices acting on the $i$th spin. The coupling strengths
\begin{align}
&F^{A}_{ij}=-2w_{ij}^2,\qquad w_{jk}=-i/\tan[(\phi_j-\phi_k)/2],\\
&F^{B}_{ij}=-2w_{ij}^2+2w_{ij}\big(\sum_{k(\neq i)}w_{ik}-\sum_{k(\neq j)}w_{jk}\big),\nonumber\\
&F^{C}=(N-2)(10N-6-N^2)/6
\nonumber\\
&\hspace{14mm}-\frac{1}{2}\sum_{i\neq j}w_{ij}^2
+(N-1)(N-2)\sum_{i}S^z_i, \nonumber
\end{align}
depend on the positions $e^{i\phi_j}$ of the spins on the unit circle. We introduce disorder into the model by choosing
\begin{equation}\label{zdisorder}
\phi_{f(j)} = 2\pi (j+ \alpha_{j})/N, \qquad j=1,2,\ldots,N,
\end{equation}
where $\alpha_{j}$ is a random number chosen with constant probability density in the interval $[-\case{\delta}{2},\case{\delta}{2}]$ and $\delta\in [0,N]$ is the disorder strength. We choose the indices $f(j)\in\{1,2,\ldots,N\}$ such that the spins are always numbered in ascending order when going around the circle. For the clean case, $\delta=0$, the spins are uniformly distributed on the circle, and for maximal disorder, $\delta=N$, all the spins may be anywhere on the circle. For $\delta\geq 1$, neighboring spins can be arbitrarily close.

\begin{figure}
\includegraphics[width=\columnwidth]{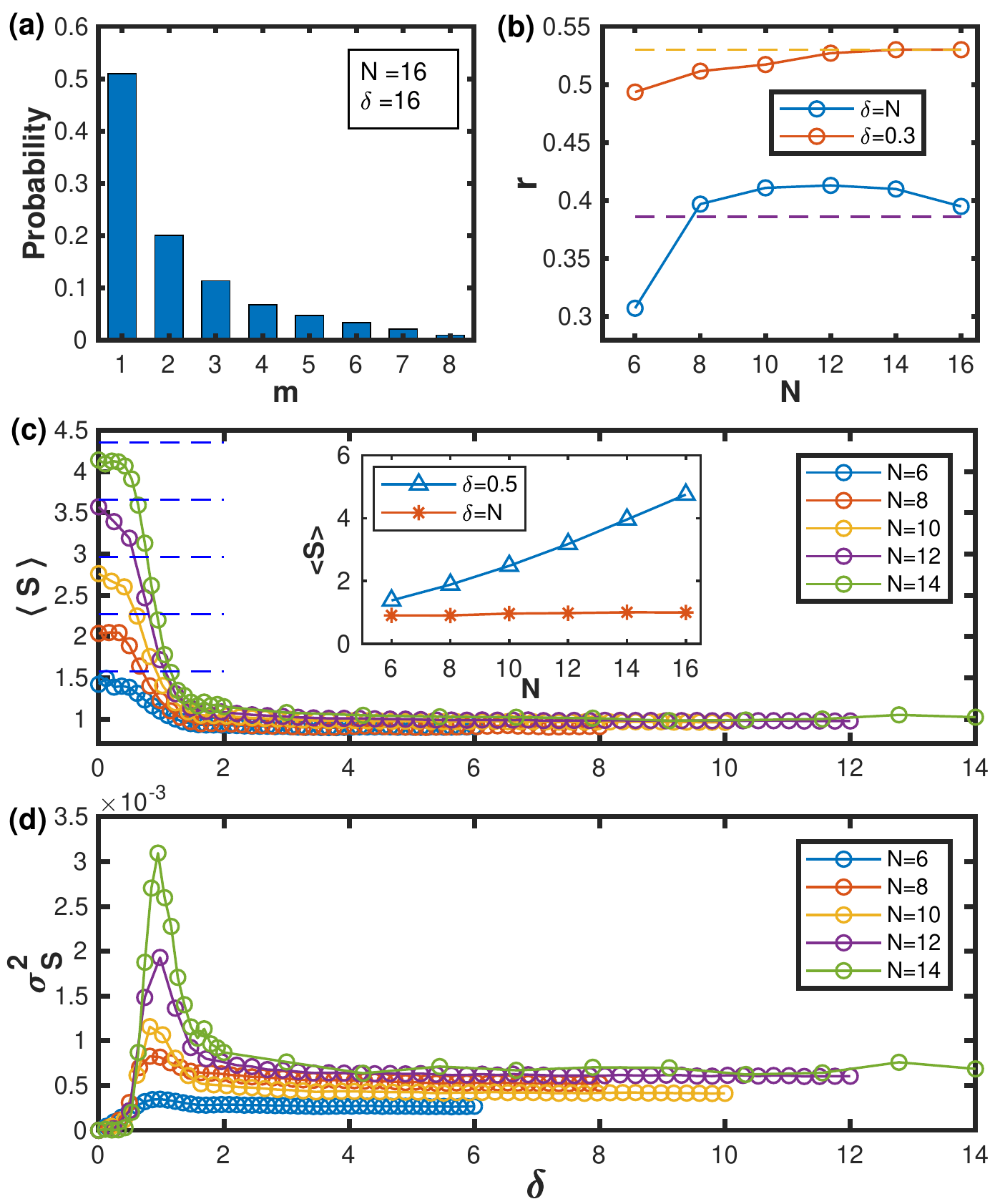}
\caption{(a) Probability that a given spin in the chain interacts most strongly with the $m$th nearest neighbor. (b) The adjacent gap ratio (averaged over $10^4$ disorder realizations and shown as a function of system size) is close to the Gaussian orthogonal ensemble for weak disorder and close to the Poisson distribution, indicating MBL, for strong disorder. (c) The transition to the MBL phase is also seen in the entanglement entropy $S$ of half of the chain for the state closest to the middle of the spectrum averaged over $10^5$ disorder realizations as a function of the disorder strength $\delta$ for different system sizes $N$. The blue dashed lines indicate the thermal value $[N\ln(2)-1]/2$ of the entropy. The inset shows that the mean entropy follows a volume law for weak disorder and an area law for strong disorder. (d) The variance $\sigma^2$ of the entanglement entropy computed from the same set of data shows a peak at the transition point.}\label{fig:entr}
\end{figure}

The Hamiltonian \eqref{ham} is fully connected, meaning that every spin interacts with every other spin. If we interpret a spin up as a particle and a spin down as an empty site, the first term in the Hamiltonian is a hopping term. The coefficient $F_{ij}^A$ decreases rapidly with distance between the spins, following a one-over-distance-squared behavior for short distances. The coefficient $F_{ij}^B$ of the interaction term has a more complicated behavior. The maximum interaction strength can occur at different distances, and Fig.\ \ref{fig:entr}(a) shows the probability that a given spin interacts most strongly with the $m$th neighbor on either the left or the right side. In the rest of this article, we restrict ourselves to the zero magnetization sector $\sum_{i} S^z_i=0$.

It was shown analytically in \cite{tu} that the state
\begin{multline}\label{psiLL}
|\psi_{0} \rangle =
\sum_{s_1,\ldots,s_N}
\delta_s\prod_k\chi_k\,
\prod_{i<j} \{\sin[(\phi_i-\phi_j)/2]\}^{(s_is_j-1)/2}\\
\times|s_1,\ldots,s_N\rangle
\end{multline}
is an exact zero energy eigenstate of \eqref{ham}. (See the supplemental material \footnote{See the supplemental material.} for the precise connection between \eqref{ham} and the Hamiltonian in \cite{tu}.) Here, $s_i=\pm1$ is twice the $z$ component of the $i$th spin, $\chi_k=e^{i\pi(k-1)(s_k+1)/2}$, and $\delta_s=1$ for $\sum_n s_n=0$ and $\delta_s=0$ otherwise. We find numerically that all eigenvalues of \eqref{ham} in the zero magnetization sector are nonnegative, also when disorder is added, so we conclude that \eqref{psiLL} is a ground state in the zero magnetization sector.

The state \eqref{psiLL} is a spin singlet \cite{Cirac}, and for $\delta=0$ it coincides with the ground state of the Haldane-Shastry model \cite{HSmodelH,shast}. We shall hence refer to the state with $\delta\neq0$ as the disordered Haldane-Shastry state. The Haldane-Shastry Hamiltonian has SU(2) symmetry, but the Hamiltonian \eqref{ham} does not. Instead, in the zero magnetization sector, it has only a global $\Z_2$ spin flip symmetry generated by $\prod^{N}_{i=1}S^x_i$.

\textit{MBL and glassiness}---We show that the highly excited states form an MBL spin glass for $\delta\agt 1$.
We first investigate the entanglement entropy $S=-\mathrm{Tr}[\rho\ln(\rho)]$ averaged over disorder realizations for states in the middle of the spectrum \cite{kjall}, where $\rho$ is the reduced density operator for half of the chain. The mean entanglement entropy and the variance of the distribution as a function of the disorder parameter $\delta$ (see Fig.\ \ref{fig:entr}(c-d)) show a phase transition at $\delta \approx 1$. The variance close to the phase transition point is large, which suggests that the transition is continuous. On the left hand side of the transition, the mean entanglement entropy scales with the system size $N$ and is bounded by the thermodynamic entropy, while on the right hand side of the transition it displays area law behaviour indicating MBL (see the inset in Fig.\ \ref{fig:entr}(c)).

The level spacing statistics is another diagnostic of MBL. We see that the energy spectrum at strong disorder ($\delta=N$) has pairs of eigenvalues that are almost degenerate and have opposite parity with respect to the global $\Z_2$ symmetry of the Hamiltonian. We hence compute the adjacent gap ratio \cite{Huse} for different system sizes by restricting ourselves to one of the $\Z_2$ sectors. Figure \ref{fig:entr}(b) shows that the disorder averaged gap ratio $r$ converges towards the Poisson distribution ($r\approx0.386$) for strong disorder and towards the Gaussian orthogonal ensemble ($r\approx0.53$) for weak disorder. This confirms that the system is indeed MBL for strong disorder.

\begin{figure}
\includegraphics[width=\columnwidth]{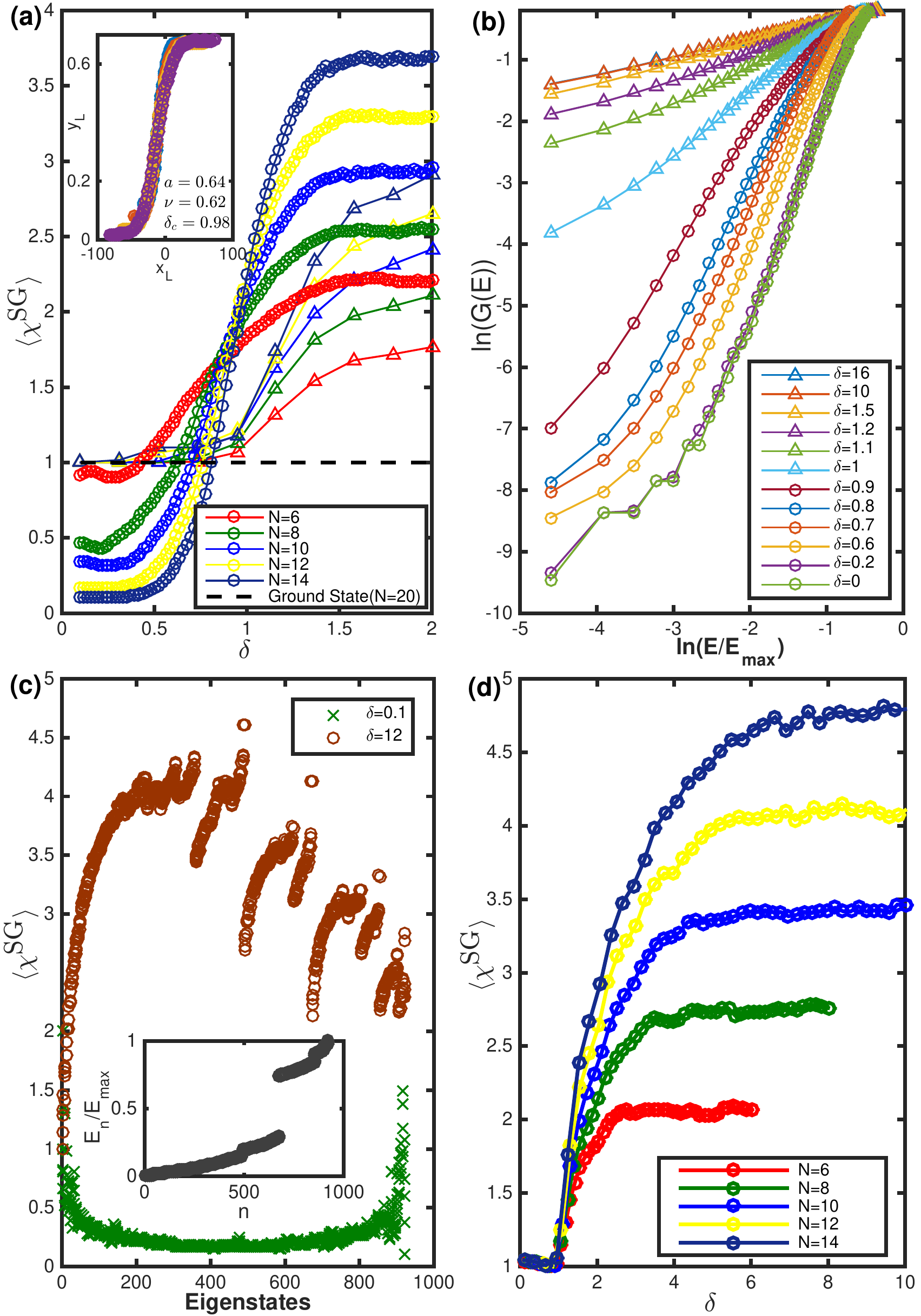}
\caption{(a) Disorder averaged spin glass order parameter $\langle\chi^{\textrm{SG}}\rangle$ for a state in the middle of the spectrum ($E/E_{\textrm{max}}=0.5$, circles), a low energy state ($E/E_{\textrm{max}}=0.01$, triangles), and the ground state ($E=0$, dashed line) as a function of the disorder strength $\delta$. The excited states show glassiness for strong disorder, and the ground state is not glassy. A finite size scaling collapse (inset) for the state in the middle of the spectrum gives the phase transition point $\delta_c\approx0.98$. (b) The integral of the density of states for $16$ spins averaged over disorder shows a transition at $\delta\approx1$. (c) $\langle\chi^{\textrm{SG}}\rangle$ for all eigenstates for a system with $12$ spins and weak ($\delta=0.1$) or strong ($\delta=12$) disorder. The states close to the ground state have different values of $\langle\chi^{\textrm{SG}}\rangle$ compared to the states in the middle of the spectrum. The inset shows that the disorder averaged energy spectrum (normalised) at $\delta=12$ has gaps, and these coincide with the jumps in $\langle\chi^{\textrm{SG}}\rangle$. (d) When we destroy the emergent SU(2) symmetry by perturbing the Hamiltonian (we modify $F^A_{ij}$ to $-1.9w^2_{ij}$), the ground state becomes glassy for $\delta\agt 1$. In all cases, the number of disorder realizations is $10^4$.}\label{fig:EAOP}
\end{figure}

The fact that pairs of almost degenerate states with opposite parity appear in the spectrum suggests that there is spin glass order in the excited states \cite{kjall,order}. In an eigenstate $|\psi_n\rangle$ this can be identified by the divergence of an Edwards-Anderson order parameter \cite{kjall,Edw}
\begin{equation}\label{EAdef}
\chi^{\textrm{SG}}=\frac{1}{N}\sum_{i\neq j}^{N}\langle \psi_n | \sigma^{z}_{i}\sigma^{z}_{j}| \psi_n\rangle^2.
\end{equation}
For eigenstates in the middle of the spectrum, we find (see Fig.\ \ref{fig:EAOP}(a)) that there is glassiness for strong disorder ($\langle\chi^{\textrm{SG}}\rangle$ increases with system size), but not for weak disorder ($\langle\chi^{\textrm{SG}}\rangle$ approaches zero with increasing system size). We perform a finite size scaling analysis to get an estimate of the critical disorder strength. The scaling parameters are given in the inset where we define the scaling function as $x_L=(\delta-\delta_c)N^{\frac{1}{\nu}}$ and $y_L=\langle\chi^{\textrm{SG}}\rangle/N^{a}$. Glassiness sets in at around the same disorder strength ($\delta_c\approx 1$) as MBL.

The transition around $\delta=1$ is also visible in the disorder averaged integral of the density of states $G(E)=\int^{E}_{0}\rho(E)dE/d_H$ plotted as a function of energy $E/E_{\textrm{max}}$ in Fig.\ \ref{fig:EAOP}(b), where $d_H$ is the total number of states in the Hilbert space and $E_{\textrm{max}}$ is the highest energy in the spectrum.

\textit{Ground state}---The low energy physics of the Haldane-Shastry model is described by Luttinger liquid theory. For strong disorder, we show that various properties of the ground state remain the same rather than reflecting a phase transition to, e.g., a random singlet phase or a glassy phase.

\begin{figure}
\includegraphics[width=\columnwidth]{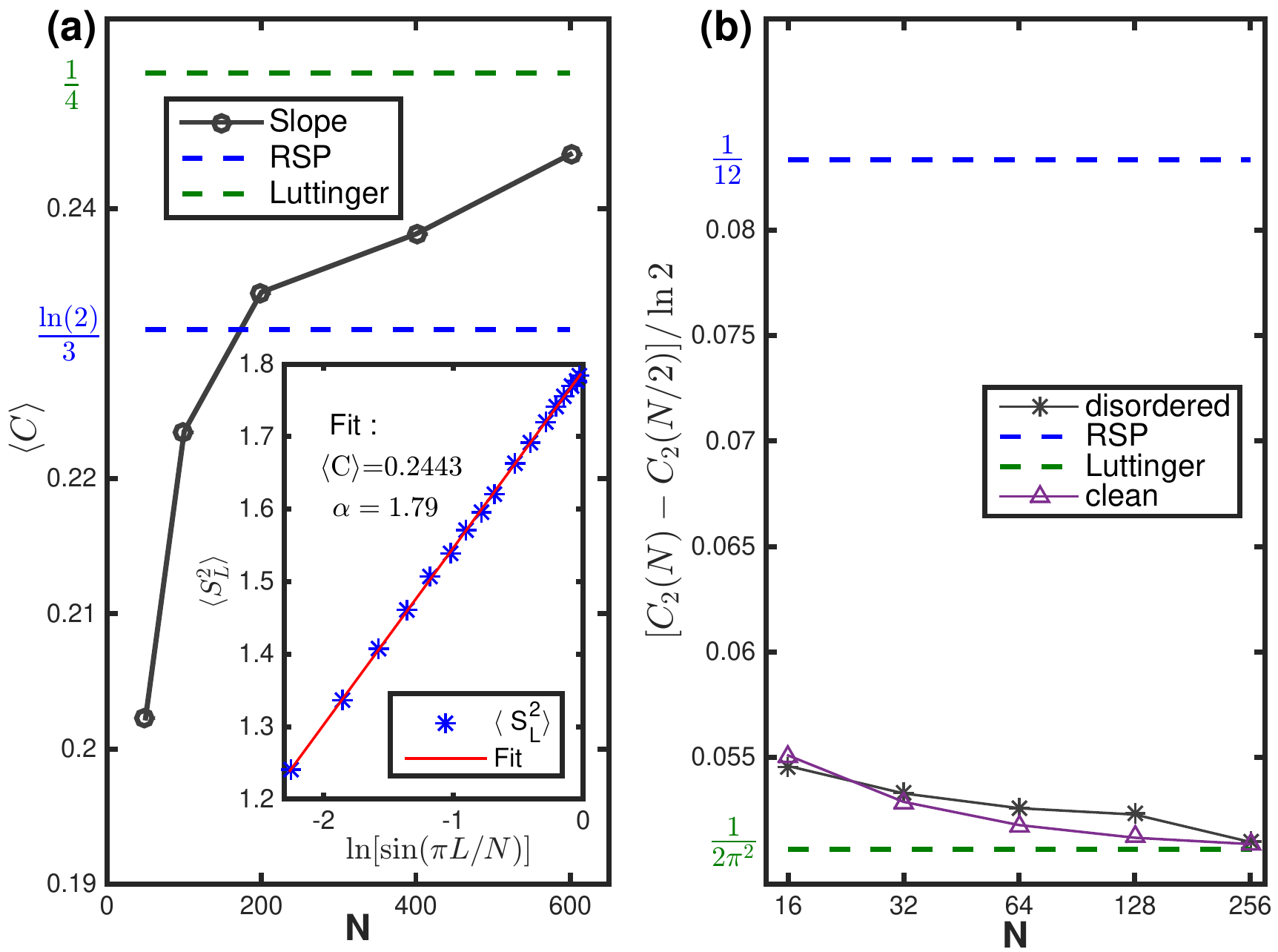}
\caption{(a) The Renyi entropy of the ground state (plotted for $N=600$ and $\delta=0.75$ in the inset) follows the logarithmic relation \eqref{entropy}. The main plot shows the coefficient $C$ as a function of the system size for $\delta=0.75$. The results might suggest a Luttinger liquid, but the method is not accurate enough to make clear conclusions. Averaging is done over $10^3$ disorder realizations, and the error occurring from the Monte Carlo simulations and disorder averaging is of order $10^{-4}$. (b) The coefficient $\xi$ in the second cumulant $C_2$ computed for different systems divided into two halves is close to the value for the Luttinger liquid both for the clean ($\delta=0$) and the disordered ($\delta=4$) Haldane-Shastry state. Each data point is averaged over $10^5$ disorder realizations.}
\label{fig:GND}
\end{figure}

The disordered Haldane-Shastry state has been studied previously for weak disorder \cite{Cirac,stephan2016full}. In \cite{Cirac}, the Renyi entropy was investigated. For critical systems it is known that the Renyi entropy of order two shows a universal behavior given by \cite{Christo,Vidal,Cala}
\begin{equation}\label{entropy}
S^2_L=C\ln\left[\sin(\pi L/N)\right]+\alpha,
\end{equation}
where $L$ is the number of spins in the considered subsystem, and $C$ is a universal constant that takes the value $1/4$ for the Luttinger liquid and $\ln(2)/3$ for the random singlet phase \cite{Fisher,Laflorencie,Shu}. Monte Carlo simulations for $\delta=0.1$, $\delta=0.5$, and $\delta=0.75$ in \cite{Cirac} showed that $C$ might be closer to $\ln(2)/3$ than to $1/4$ for $\delta=0.5$ and $\delta=0.75$. We have redone the computation for $\delta=0.75$ in Fig.\ \ref{fig:GND}, and the results suggest that $C$ might rather go to $1/4$ for large system sizes. The differences between \cite{Cirac} and our results could be due to that only $L$ values close to $N/2$ were used for the fitting in \cite{Cirac}. The main conclusion from the computations, however, is that the uncertainty in determining $C$ due to, e.g., finite size effects and ambiguity in the fitting procedure is not small compared to the difference between $1/4$ and $\ln(2)/3$. The conclusions may not be reliable, and one should also consider other diagnostics.

In \cite{stephan2016full}, the second cumulant
\begin{equation}\label{cumulant}
C_{2}(N/2)=\langle M^2\rangle-\langle M\rangle^2, \quad M\equiv\sum_{i=1}^{N/2}S_{i}^{z},
\end{equation}
of the total magnetization $M$ of half of the system was observed to show a Luttinger liquid behavior at $\delta=1$. Specifically, the second cumulant is known to diverge logarithmically with system size, $C_{2}(N/2)\sim \xi \ln(N/2)+\textrm{constant}$ for large $N$, with different coefficients for the Luttinger liquid ($\xi=1/(2\pi^2)$) and for the random singlet phase ($\xi=1/12$), and it was shown that $[C_{2}(N)-C_{2}(N/2)]/\ln(2)\approx\xi$ approaches the value for the Luttinger liquid for large system sizes. Figure \ref{fig:GND} shows that the same is true for $\delta=4$. This shows that the ground state retains its Luttinger liquid behavior also for a disorder strength for which the highly excited states are MBL. Note that relatively large system sizes can be reached in these computations because the two point correlations can be obtained by solving a set of linear equations derived in \cite{Nielsen_2011}.

Finally, we consider the spin glass order parameter shown in Fig.\ \ref{fig:EAOP}(a). Here, the ground state yet again stands out, with no indication of a phase transition. In fact, for the ground state,  $\chi^{\textrm{SG}}=1$ is constant, irrespective of disorder strength and system size.

\textit{Low-lying excited states}---The observation that the highly excited states undergo a transition to MBL, while the ground state does not undergo a transition, naturally raises the question, whether it is only the ground state that is special, or the ground state properties are to some extent inherited to the low-lying excitations. By studying the spin glass order parameter, we find that a small number of low-lying excitations behave differently, but as soon as we consider a finite energy density, the states appear to show glassiness for strong disorder.

In Fig.\ \ref{fig:EAOP}(a), we show data for $\langle\chi^{\textrm{SG}}\rangle$ for the case where we choose the state in the spectrum that is closest to $E/E_{\textrm{max}}=0.01$ in every disorder realization. Even for this low value of the energy density, glassiness is still observed for strong disorder. A more detailed view for $12$ spins is given in Fig.\ \ref{fig:EAOP}(c), where we plot $\langle\chi^{\textrm{SG}}\rangle$ for all states in the spectrum for weak and strong disorder. For all the highly excited states, there is a large increase in $\langle\chi^{\textrm{SG}}\rangle$ when going from weak to strong disorder, which shows the transition into the glassy phase. For the ground state, there is no change as $\chi^{\textrm{SG}}=1$. A few states close to the ground state show an intermediate behavior and have particularly low values of $\langle\chi^{\textrm{SG}}\rangle$ for strong disorder. The inset shows the disorder averaged spectrum, and we note that the sudden jumps observed in $\langle\chi^{\textrm{SG}}\rangle$ coincide with gaps in the spectrum.

\textit{Symmetry}---Finally, we show that the special behavior of the ground state disappears together with the emergent SU(2) symmetry. To do so, we slightly modify the hopping strengths $F^A_{ij}$ to $-1.9w_{ij}^2$. This preserves the $\Z_2$ symmetry of the Hamiltonian, but not the emergent SU(2) symmetry of the ground state. Figure \ref{fig:EAOP}(d) shows that the ground state is now glassy for strong disorder. If instead we add a small amount of the Haldane-Shastry Hamiltonian, the ground state is unaltered, and the $\Z_2$ symmetry of the Hamiltonian is preserved. In this case, the spin glass order parameter behaves similarly to the results in Fig.\ \ref{fig:EAOP}(a).

\textit{Conclusions}---We have constructed a new type of MBL model, in which an emergent symmetry protects the ground state from MBL. While states at a finite energy density show MBL for sufficiently strong disorder, the ground state remains critical. It seems likely that the observed behavior is a general mechanism to protect states from MBL, and it would be interesting to search for a similar behavior in other models.

The model has the unexpected property that the disorder averaged energy spectrum has gaps. The background for this is not understood and would be interesting to investigate further.

While we do find glassiness to be present in all excited states already at very low energy densities, we find a different behavior for the states adjacent to the ground state. The possibility of a `critical regime' with a diverging localization length warrants further study here.

If the delocalized state should turn out to be genuinely isolated in a sea of localized ones, this would have the flavour of an `inverted scar state', i.e.\ a state with above-area-law entanglement in a sea of area law entangled states; it is the converse of what is found in the celebrated many-body scars \cite{scars}.

At any rate, it would be interesting to consider the scope of constructing models with multiple states in the spectrum -- including at finite energy density -- being protected from localization by an emergent symmetry.

Another interesting direction for further investigations is to study the transport properties of our model, in particular in a regime where the ground state and some of the lower lying excited states are populated.

\begin{acknowledgments}
\textit{Acknowledgements}---We thank Giuseppe De Tomasi and Ivan Khaymovich for discussions. This work was in part supported by the Deutsche Forschungsgemeinschaft under grants SFB 1143 (project-id 247310070) and the cluster of excellence ct.qmat (EXC 2147, project-id 39085490).
\end{acknowledgments}

\begin{widetext}
\end{widetext}

\appendix

\textit{Supplemental material}---Our starting point is the Hamiltonian
\begin{equation}\label{eqham}
H=\sum_{i}\Lambda^{\dagger}_{i}\Lambda_{i}-2\sum_{i}\Gamma^{\dagger}_{i}\Gamma_{i}
\end{equation}
for hardcore bosons on a lattice introduced in \cite{tu}. Here,
\begin{align}
&\Lambda_{i}=\sum_{j(\neq i)}w_{ij}[c_{j}-c_{i}(2n_{j}-1)],\label{eqlambda}\\
&\Gamma_{i}=\sum_{j(\neq i)}w_{ij}c_{i}c_{j}\label{eqgamma},
\end{align}
where $c_j$ is the operator that annihilates a hardcore boson on site $j$, and $n_j=c^{\dag}_jc_j$. It can be shown \cite{tu} that both $\Lambda_i$ and $\Gamma_i$ annihilate the state in Eq.\ (4) in the main text.

Inserting \eqref{eqlambda} and \eqref{eqgamma} into \eqref{eqham}, we obtain
\begin{equation}\label{ham}
H=\sum_{i\neq j}(F^{A}_{ij} \, c_{i}^{\dagger}c_{j} +F_{ij}^{B}\, n_{i}n_{j}) +\sum_{i}F^{C}_{i}\, n_i+F^{D},
\end{equation}
where the coupling coefficients are given by
\begin{equation}
\begin{split}
&F^{A}_{ij}=-2w_{ij}^2,\\
&F^{B}_{ij}=2w_{ij}^2+4\sum_{l(\neq j\neq i)}w_{ij}w_{il},\\
&F^{C}_{i}=-2\sum_{j(\neq i)}w_{ij}^2-\sum_{k,l(\neq i)}w_{ik}w_{il},\\
&F^{D}=\frac{-N(N-2)(N-4)}{6}.
\end{split}
\end{equation}

One arrives at the spin version of the Hamiltonian described in Eq.\ (1) of the main text by introducing the transformation
\begin{equation}
S_i^{+}=c^{\dagger}_i \;\;\;\;\;S_i^{-}=c_i \;\;\;\;\;S^{z}_i=c^{\dagger}_ic_i-1/2,
\end{equation}
where $S_{i}^{\pm}=S^{x}_i\pm iS^{y}_{i}$.

\end{document}